\documentclass[11pt]{article}
\usepackage{arxiv}
\usepackage[utf8]{inputenc}
\usepackage[T1]{fontenc}
\usepackage{hyperref}
\usepackage{url}
\usepackage{booktabs}
\usepackage{amsfonts}
\usepackage{nicefrac}
\usepackage{microtype}
\usepackage{graphicx}
\usepackage[numbers]{natbib}
\usepackage{doi}
\usepackage{xcolor}
\usepackage[ruled]{algorithm2e}
\usepackage{braket}
\usepackage{amsmath}
\usepackage[amsmath,thmmarks]{ntheorem}
\usepackage{cleveref}
\usepackage[hang,small,bf]{caption}
\usepackage[subrefformat=parens]{subcaption}
\usepackage{url}
\usepackage{tikz}
\usepackage{qcircuit}
\usepackage{todonotes}
\newtheorem{theorem}{Theorem}

\title{The Role of Entanglement in Quantum-Relaxation Based Optimization Algorithms}

\author{ {\hspace{1mm}Kosei Teramoto}\\
	Dept. of Computer Science, The Univ. of Tokyo\\
	\texttt{teramoto@is.s.u-tokyo.ac.jp}\\
	\And
	{\hspace{1mm}Rudy Raymond}\\
	IBM Quantum, IBM Japan\\
	Dept. of Computer Science, The Univ. of Tokyo\\
	Quantum Computing Center, Keio University\\
	\texttt{rudyhar@jp.ibm.com}\\
	\And
	{\hspace{1mm}Hiroshi Imai}\\
	Dept. of Computer Science, The Univ. of Tokyo\\
	\texttt{imai@is.s.u-tokyo.ac.jp}\\
}

\renewcommand{\headeright}{~}
\renewcommand{\undertitle}{~}
\renewcommand{\shorttitle}{The Role of Entanglement in Quantum-Relaxation Based Optimization Algorithms}

\hypersetup{
pdftitle={The Role of Entanglement in Quantum-Relaxation Based Optimization Algorithms},
pdfsubject={Quantum Physics (quant-ph)},
pdfauthor={Kosei Teramoto, Rudy Raymond, Hiroshi Imai},
pdfkeywords={Quantum Relaxation, Combinatorial Optimization, Variational Quantum Algorithms, Benchmarks},
}

\begin{document}
\maketitle

\begin{abstract}
Quantum Random Access Optimizer (QRAO) is a quantum-relaxation based optimization algorithm proposed by Fuller et al. that utilizes Quantum Random Access Code (QRAC) to encode multiple variables of binary optimization in a single qubit. Differing from standard quantum optimizers such as QAOA, it utilizes the eigenstates of local quantum Hamiltonians that are not diagonal in the computational basis. There are indications that quantum entanglement may not be needed to solve binary optimization problems with standard quantum optimizers because their maximal eigenstates of diagonal Hamiltonians include classical states.
In this study, while quantumness does not always improve the performance of quantum relaxations, we observed that there exist some instances in which quantum relaxation succeeds to find optimal solutions with the help of quantumness.
Our results suggest that QRAO not only can scale the instances of binary optimization problems solvable with limited quantum computers but also can benefit from quantum entanglement.
\end{abstract}

\section{Introduction}
Solving optimization problems is one of the most important tasks for which quantum computation is expected to be useful.
Various quantum algorithms have been devised for optimization problems.
Among them, QAOA (Quantum Approximate Optimization Algorithm) \cite{farhi2014quantum}, which generalizes adiabatic quantum computation \cite{farhi2000quantum} to optimization problems proposed by Farhi, Goldstone, and Gutmann, and VQE (Variational Quantum Eigensolver) \cite{peruzzo2014variational} proposed by Peruzzo, et al. are well known. They have been applied to various NP-hard problems (i.e. maximum cut problem and graph coloring problem).
QAOA and VQE are classical-quantum hybrid algorithms for near-term devices where only shallow quantum circuits are realized.
However, they have some issues.

The first issue is scalability.
Because QAOA and VQE encode one classical bit into one qubit, a quantum device that has $n$ qubits is needed to solve an instance of optimization problems with size $n$.
Current quantum devices such as the IBM Quantum devices are small in scale and contain noise that is not fault-tolerant, so as the size of the circuit increases, the effect of the noise becomes too large to make the calculation results meaningful.
Therefore, in the near future, it will be necessary to devise smaller circuits to solve larger problems.
The second issue is that we do not know if the \textit{quantumness} of constant-depth QAOA and VQE can give rise to better results than the classical optimization algorithm.
This is because there are experimental results that show no difference in solving various NP-hard problems in variational circuits with and without entanglement in the ansatz \cite{nannicini2019performance}\cite{diez2021quantum}, while there are instances when constant-depth QAOA are worse than the classical counterpart~\cite{bravyi2020obstacles}. Therefore, it is possible that hybrid optimization algorithms of the type that directly search for classical solutions, such as VQE, do not require the quantum mechanical property of quantum entanglement to search for solutions.
In other words, these algorithms may not need to be run on a quantum computer in the first place.
This is a problem that is related to the very significance of solving optimization problems on a quantum computer.\\

Recently, a new classical-quantum hybrid optimization algorithm, QRAO (Quantum Random Access Optimization) \cite{fuller2021approximate}, was proposed by Fuller et al. to address the above problem. QRAO is based on QRA codes~\cite{ambainis2002dense} that can encode up to $4^n-1$ binary variables into $n$ qubits~\cite{INRY07}. 
Specifically, the QRAO encodes multiple classical bits (less than or equal to three) into one qubit using the (3,1)-QRA code~\cite{HINRY06}.
Where QAOA and VQE, for example, use only the Pauli ${Z}$ operator to encode classical information into quantum information, QRAO also uses the Pauli ${X}$ and Pauli ${Y}$ operators to encode more classical information into quantum information. This way Fuller et al.~\cite{fuller2021approximate} were able to perform experiments with QRAO on superconducting quantum devices to solve the largest instances of the maximum cut problem.
We note that there exist different heuristics from QRAO which only use a logarithmic number of qubits to the size of the problem instance~\cite{Tan2021qubitefficient,ranvcic2021exponentially,winderl2022comparative}.
These heuristics use problem formulation with graph Laplacian matrices and are applied to various NP-hard problems~\cite{chatterjee2023solving}.
Though the space efficiency of them are significant, they do not have a theoretical performance guarantee like QRAO described in \Cref{thm:31ratio}.
Also, since QRAO searches for quantum states that correspond to solutions to the relaxation problem rather than classical solutions, the quantum state that is eventually discovered is an entangled state, due to the use of Pauli $X$ and $Y$ in the quantum relaxation, that cannot be directly interpreted as a classical solution.
To obtain a classical solution, \textit{quantum state rounding} of the relaxed solution must be performed. 
Therefore, compared to standard VQE methods, the ansatz of QRAO may inherently require quantum entanglement.
In other words, VQE for QRAO are inherently different from classical algorithms and may benefit from quantum mechanical properties such as entanglement.\\

In this study, we hypothesized that introducing quantum entanglement into ansatz when solving optimization problems with quantum relaxation such as QRAO would improve the results, and verified this hypothesis by extensive experiments of solving maximum cut problems on simulators. 
As a result, we find some instances for which the quantum relaxation without entanglement layers fails to find an optimal solution while that with entanglement layers succeeds in getting an optimal solution.

\section{Quantum-Relaxation Based Optimization Algorithm}
The following is based on~\cite{fuller2021approximate}.
We explain the quantum-relaxation based optimization algorithm by using the MaxCut problem.
Let $G=(V(G),E(G))$ be a graph with $|V(G)|$ vertices and $|E(G)|$ edges.
The vertices and the edges are labeled as $\{v_i\}$ and $\{e_{i,j}\}$ where $i,j\in[|V(G)|]$.
Then MaxCut problem can be formulated like the following.
\begin{equation}
    \max_{x\in\{+1, -1\}^{|V(G)|}}\frac{1}{2}\sum_{e_{i,j}\in E(G)}\left(1-x_ix_j\right)
    \label{eq:MaxCut}
\end{equation}

\subsection{Generating Relaxed Hamiltonian}
The methods solving the problem directly on quantum computers such as VQE or QAOA, each classical binary variable $x_i$ is mapped to $i$-th qubit using the Pauli $Z$ operator.
Let $Z_i$ be a Pauli $Z$ operator applied to $i$-th qubit.
Then, the MaxCut problem is reduced to the problem to find a maximum eigenstate of the Hamiltonian
\begin{equation}
    H=\frac{1}{2}\sum_{e_{ij}\in E}\left(I-Z_iZ_j\right).
    \label{eq:H}
\end{equation}
Variational methods are used to search for the maximum eigenstate of $H$.
Because $H$ is a diagonal Hamiltonian, it contains the classical state (without superposition and entanglement) as the maximal eigenstates so that the found state in the algorithm can be interpreted directly as the classical solution to the MaxCut problem by just measuring it in the computational basis.

On the other hand, in the quantum-relaxation based optimization algorithms such as QRAO~\cite{fuller2021approximate}, multiple classical bits are encoded into a smaller number of qubits using QRACs.
For example, let us consider the case to use $\left(3,1,\frac{1}{2}+\frac{1}{2\sqrt{3}}\right)$-QRAC (or just $(3,1)$-QRAC)~\cite{ambainis1999dense,ambainis2002dense,HINRY06} which encodes three binary variables $x_1, x_2, x_3 \in \{0, 1\}$ into a single-qubit defined by the following equation:
\begin{equation}
    (x_1,x_2,x_3)\mapsto\rho_{x_1,x_2,x_3}:=\frac{1}{2}\left(I+\frac{1}{\sqrt{3}}((-1)^{x_1}X+(-1)^{x_2}Y+(-1)^{x_3}Z\right).
\end{equation}
The first parameter of the QRAC, $3$, is the number of classical binary variables encoded into the number of qubits designated by the second parameter $1$.
The third parameter $\frac{1}{2}+\frac{1}{2\sqrt{3}}$ corresponds to the success probability of decoding each encoded bit.
Then, three classical variables $x_1$, $x_2$, and $x_3$ are mapped to a single-qubit using the Pauli $X$, $Y$, and $Z$ operators respectively.
Compared with QAOA or VQE, QRAO has the constant-factor space complexity advantage.
We will focus on the QRAO using $(3,1)$-QRAC from here.
The goal is, as well as the typical methods, to reduce the MaxCut problem to the procedure to explore the maximum eigenstate of the Hamiltonian called \textit{relaxed Hamiltonian} $H_{relax}$.
To construct a relaxed Hamiltonian, we make the mapping from classical binary variables into qubits.
Firstly, we perform a coloring of the graph $G$.
After coloring, the vertices are partitioned into the set $\{V_c\}$ associated with the color $c\in C$.
Let $\mathrm{color(i)}$ be the color of the $i$-th vertex $v_i$.
Then, the following condition holds:
\begin{equation}
    e_{i,j}\in E(G)\implies\mathrm{color}(i)\neq\mathrm{color}(j).
\end{equation}
Next, we associate $\lceil\frac{|V_c|}{3}\rceil$ qubits for each color $c\in C$.
Now up to three vertices are assigned to a single qubit.
We greedily order these three vertices and assign the Pauli $X$, $Y$, and $Z$ respectively.
Finally, we obtained a relaxed Hamiltonian instead of the normal Hamiltonian in \Cref{eq:H} as below:
\begin{equation}
    H_{relax}:=\frac{1}{2}\sum_{e_{i,j}\in E(G)}(I-3P_iP_j),
    \label{eq:H_relax}
\end{equation}
where $P_i$ is the Pauli operator associated with the vertex $v_i$.
We explore the maximum eigenstate of $H_{relax}$ by using the variational methods.
The relaxed Hamiltonian $H_{relax}$ is no longer diagonal and for many cases it contains the non-classical states (with superposition and entanglement) as the maximal eigenstates.
It means that the found eigenstate for the relaxed Hamiltonian cannot be associated with the classical solution directly.
Because of the construction of the Hamiltonian, the found state should be a quantum state that corresponds to the relaxed solution to the MaxCut problem.
A relaxed solution means the solution of the MaxCut problem without the constraint that the solution must be a binary vector.
We denote the found eigenstate in quantum-relaxation based optimization algorithm as $\rho_{relax}$ and call it \textit{relaxed state}.

\subsection{Rounding Relaxed Quantum States}
To retrieve the classical solution for the MaxCut problem from the obtained relaxed state $\rho_{relax}$, we perform quantum state rounding algorithms.
There are two types of rounding algorithms proposed by Fuller et al.~\cite{fuller2021approximate}.

\subsubsection{Pauli Rounding Algorithm}
The first rounding algorithm is \textit{Pauli rounding} which decode the encoded three classical bit in each qubit by using the quantum measurements corresponding to the success probability $\frac{1}{2}+\frac{1}{2\sqrt{6}}$ associated with $(3,1)$-QRAC.
More precisely, we perform $X$, $Y$, and $Z$ basis measurements for all qubits with enough shots and calculate the expectation of the $\mathrm{Tr}[M(v_i)\rho_{relax}]$ denoted by $\mathrm{est}_i$ for all vertices $i\in V(G)$ where $M$ is an assignment from vertices to Pauli operators.
After that, we decode the corresponding classical binary value according to $\mathrm{sign}(est_i)$.
This procedure is equivalent to just measuring the $j$-th qubit with enough shots, taking the majority of the measurement result, and setting it to the rounded value of the corresponding classical bit.
The whole procedure is described in \Cref{alg:pauli}.
\begin{algorithm}[tb]
    \SetKwInOut{Input}{Input}\SetKwInOut{Output}{Output}\SetAlgoNoLine
    \caption{Pauli rounding algorithm}
    \label{alg:pauli}
    \Input{An oracle $\mathcal{O}_{relax}$ which prepares relaxed state $\rho_{relax}$; Number of measurement shots $S$; An assignment $M$ from vertex to Pauli operator.}
    \Output{Approximate solution $x\in\{0,1\}^{|V(G)|}$}
    Initialize approximate solution $x=(1,1,...,1)$.\\
    Prepare $\rho_{relax}$ using $\mathcal{O}_{relax}$.\\
    Measure each qubit on $X$, $Y$, and $Z$ basis with $S$ shots respectively.\\
    Calculate the estimation $\mathrm{est}_i$ of the value $Tr\left(\rho_{relax}\cdot M(v_i)\right)$ for each $v_i\in V(G)$.\\
    \For{$i\in [|V(G)|]$}{
        \eIf{$(\mathrm{est}_i=0)$}{
            Assign the value to $x_i$ uniformly at random.
        }{
            Assign the value to $x_i$ according to $\mathrm{sign}(\mathrm{est}_i)$.
        }
    }
    \Return{$x$}
\end{algorithm}
If the relaxed state can be written as the product state like $\rho_1\otimes\rho_2\otimes\cdots\rho_n$, the Pauli rounding algorithm works well.
On the other hand, if the quantum state is very entangled and cannot be written in the form $\rho_1\otimes\rho_2\otimes\cdots\rho_n$, there is no guarantee that the Pauli rounding performs well.
This is because the Pauli rounding algorithm does not consider the correlation among different qubits.

\subsubsection{Magic State Rounding Algorithm}
\begin{figure}[tb]
    \begin{tabular}{cc}
        \begin{minipage}[t]{0.45\hsize}
            \centering
            \includegraphics[height=5cm]{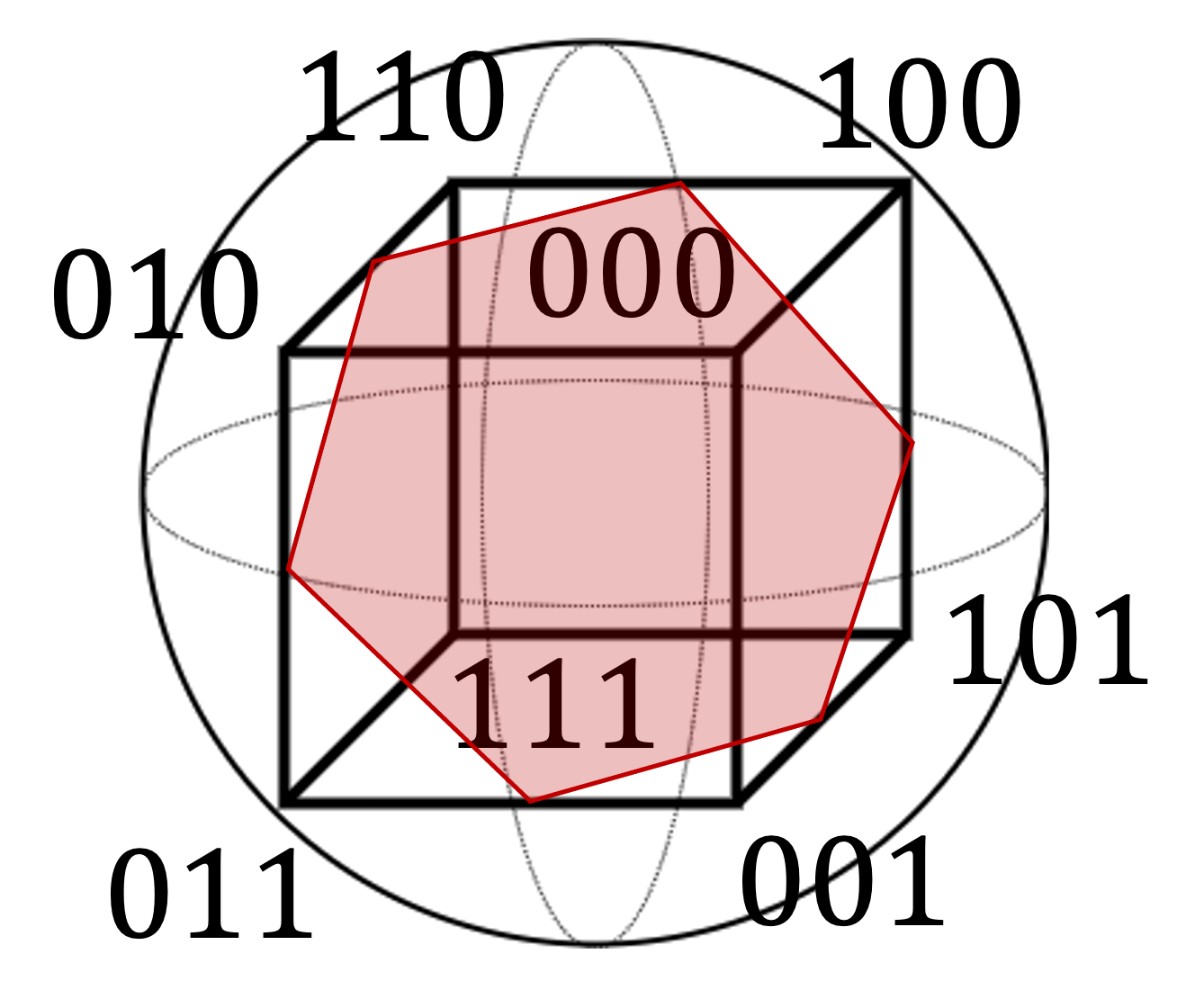}
            \subcaption{$\mu^{\pm}_1:=\frac{1}{2}\left(I\pm\frac{1}{\sqrt{3}}(X+Y+Z)\right)$}       
        \end{minipage} &
        \begin{minipage}[t]{0.45\hsize}
            \centering
            \includegraphics[height=5cm]{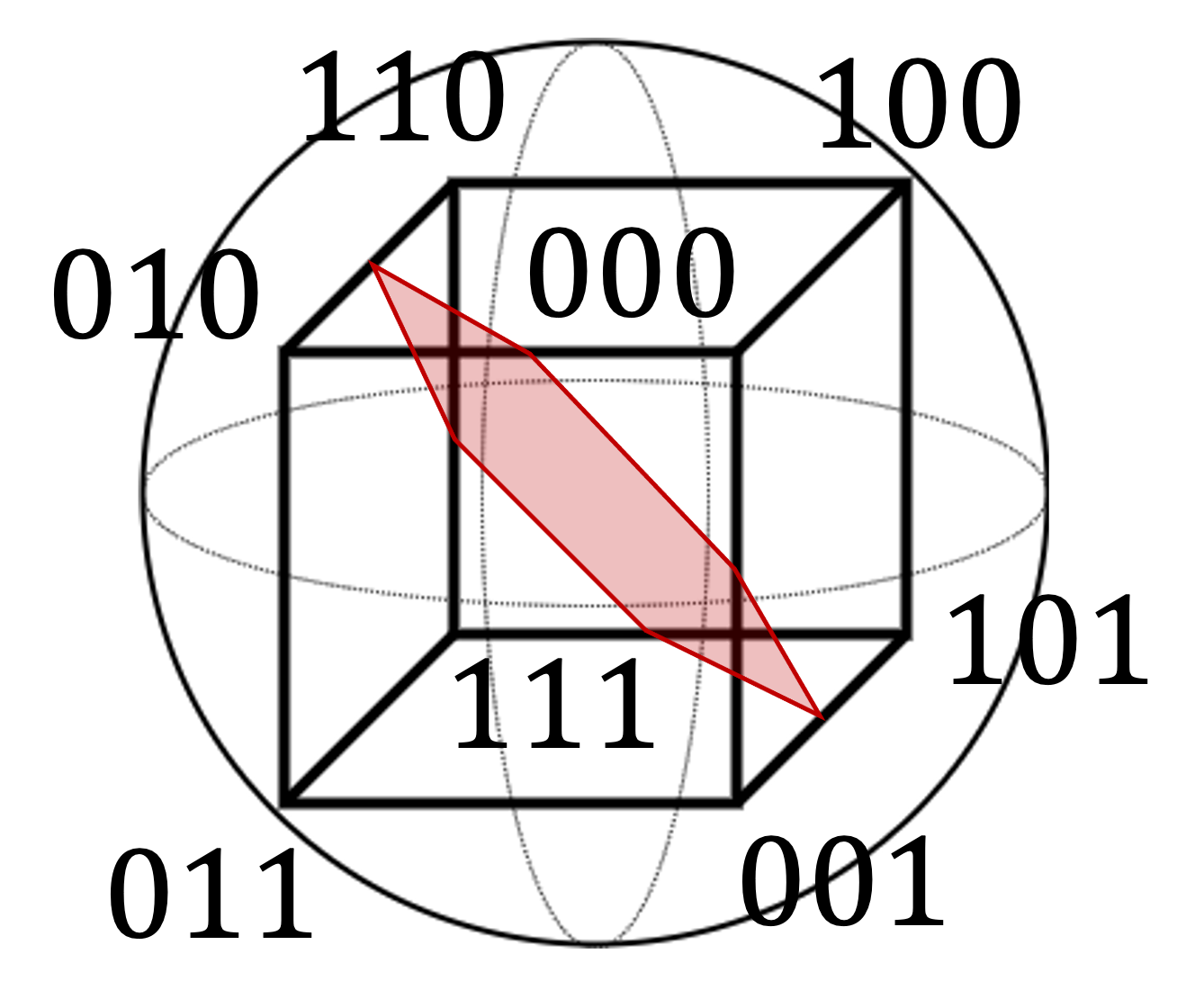}
            \subcaption{$\mu^{\pm}_2:=\frac{1}{2}\left(I\pm\frac{1}{\sqrt{3}}(X-Y-Z)\right)$}
        \end{minipage} \\
        \begin{minipage}[t]{0.45\hsize}
            \centering
            \includegraphics[height=5cm]{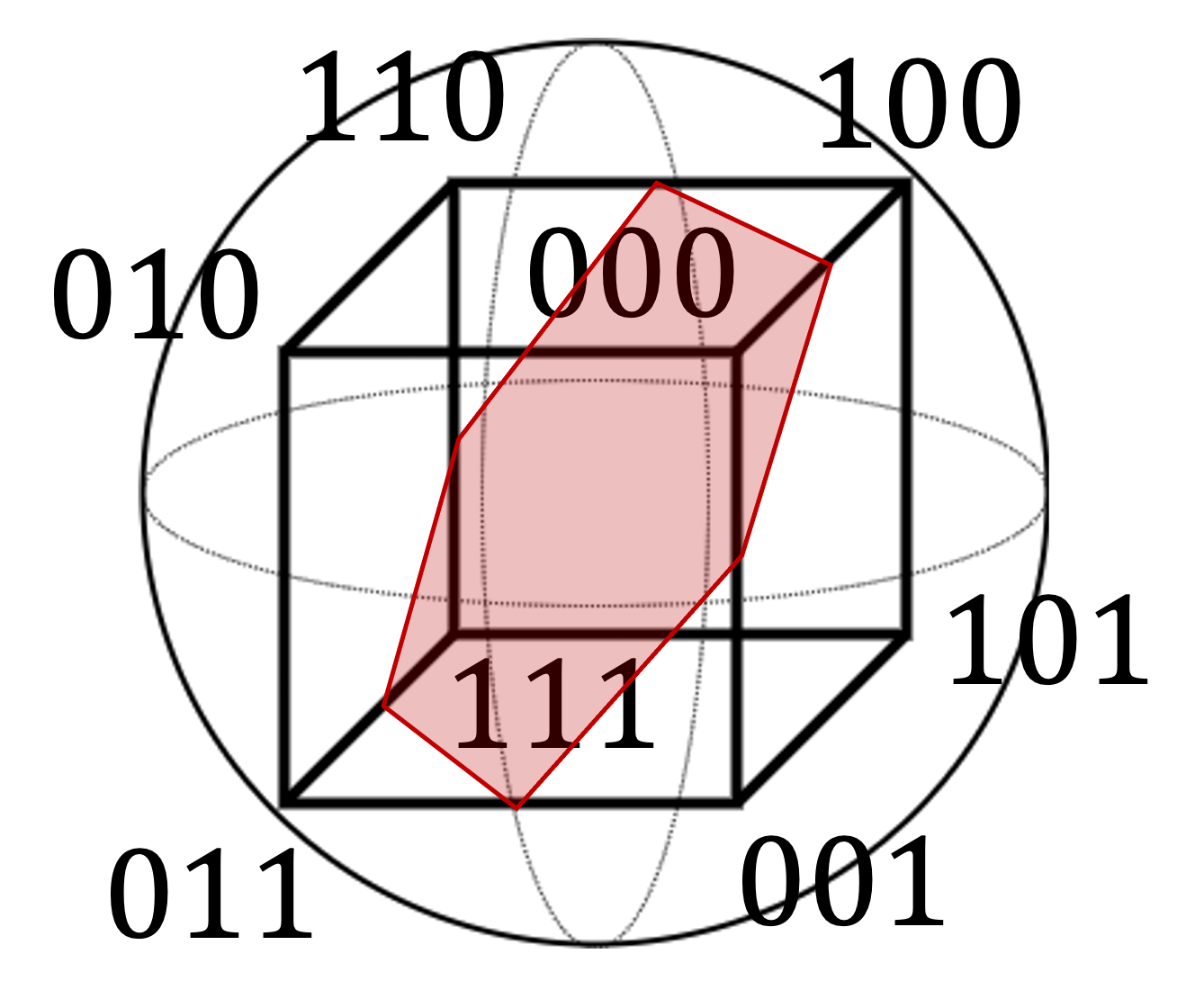}
            \subcaption{$\mu^{\pm}_3:=\frac{1}{2}\left(I\pm\frac{1}{\sqrt{3}}(-X+Y-Z)\right)$}       
        \end{minipage} &
        \begin{minipage}[t]{0.45\hsize}
            \centering
            \includegraphics[height=5cm]{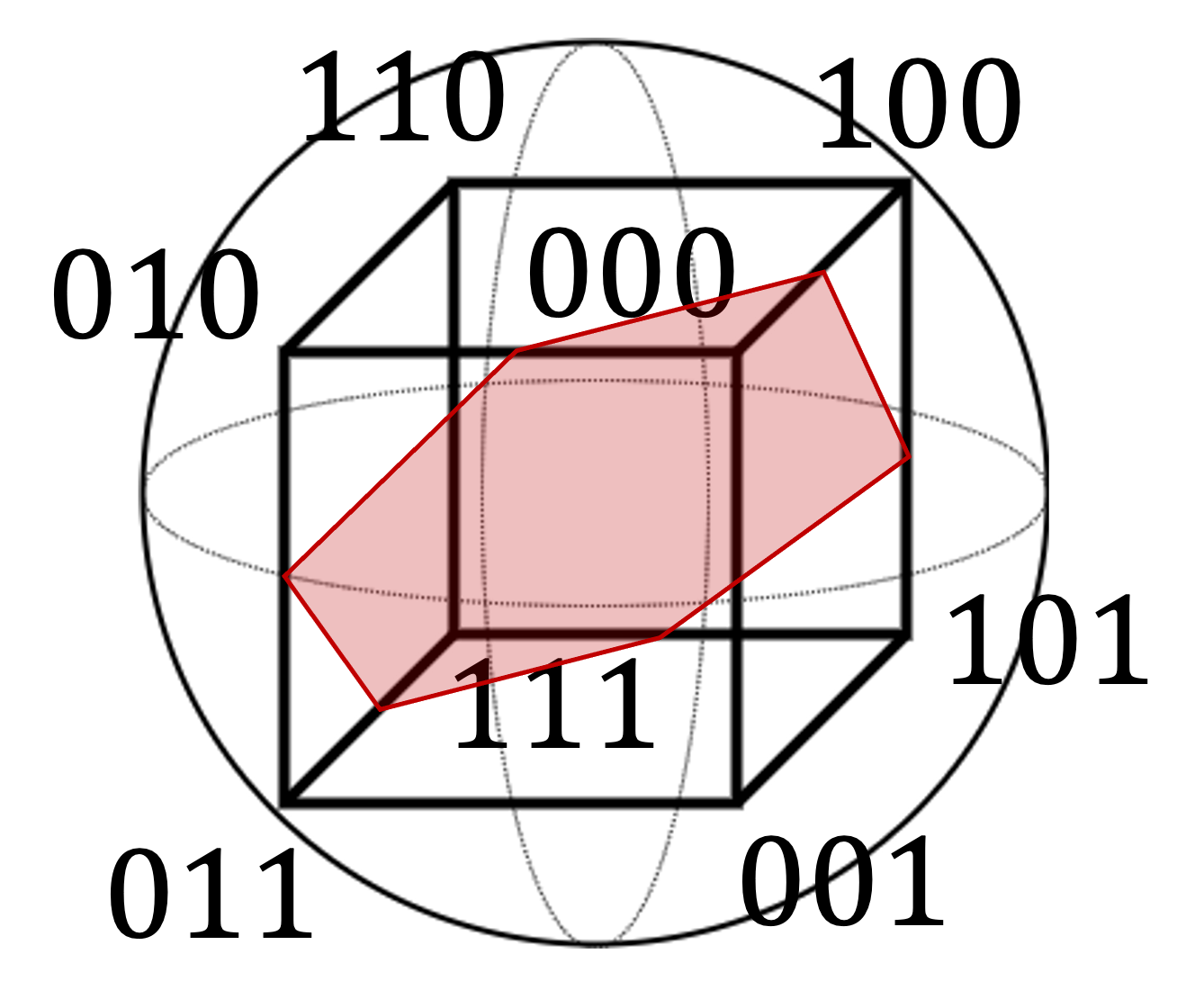}
            \subcaption{$\mu^{\pm}_4:=\frac{1}{2}\left(I\pm\frac{1}{\sqrt{3}}(-X-Y+Z)\right)$}
        \end{minipage}
    \end{tabular}
    \caption{Illustration of the quantum measurements performed in magic state rounding algorithm}
    \label{fig:magic}
\end{figure}

By using the second rounding algorithm, \textit{magic state rounding}, we can avoid the aforementioned problem and can obtain the approximation ratio bound for the MaxCut problem.
The idea of the magic state rounding algorithm is to decode three classical variables at once from a single qubit.
Consider the single-qubit magic state:
\begin{equation}
    \mu^{\pm}:=\frac{1}{2}\left(I\pm\frac{1}{\sqrt{3}}(X+Y+Z\right)
\end{equation}
and set
\begin{align*}
    \mu^{\pm}_1&:=\mu^{\pm},\\
    \mu^{\pm}_2&:=X\mu^{\pm}X=\frac{1}{2}\left(I\pm\frac{1}{\sqrt{3}}(X-Y-Z)\right),\\
    \mu^{\pm}_3&:=Y\mu^{\pm}Y=\frac{1}{2}\left(I\pm\frac{1}{\sqrt{3}}(-X+Y-Z)\right),\\
    \mu^{\pm}_4&:=Z\mu^{\pm}Z=\frac{1}{2}\left(I\pm\frac{1}{\sqrt{3}}(-X-Y+Z)\right).
\end{align*}
In the magic state rounding algorithm, one fo the measurement basis $\{\mu^+_i,\mu^-_i\}$ is selected from $i\in[4]$ for each qubit.
After choosing the bases for all qubits, then a relaxed state $\rho_{relax}$ is measured on those bases.
Three classical binary variables are decoded according to the measurement outcome for each qubit.
\Cref{fig:magic} shows the intuition of the magic state rounding algorithm.
Each measurement $\mu^{\pm}_i$ decodes one of the pair of three bits located at opposite angles on the cube (e.g. $000$ or $111$ in the case of $\mu^{\pm}_1$).
By using this simultaneous decoding of the encoded three bits, the magic state rounding algorithm extracts the solution of the MaxCut for every iteration.
The magic state rounding algorithm repeats this procedure enough times and outputs the best solution.
The whole procedure is described in \Cref{alg:magic}.
\begin{algorithm}[tb]
    \SetKwInOut{Input}{Input}\SetKwInOut{Output}{Output}\SetAlgoNoLine
    \caption{Magic state rounding algorithm}
    \label{alg:magic}
    \Input{An oracle $\mathcal{O}_{relax}$ which prepares relaxed state $\rho_{relax}$; Number of measurement shots $S$.}
    \Output{Approximate solution $x\in\{0,1\}^{|V(G)|}$}
    Initialize approximate solution $x=(1,1,...,1)$.\\
    \For{$s\in[S]$}{
        Prepare $\rho_{relax}$ using $\mathcal{O}_{relax}$.\\
        Randomly and independently choose measurement basis for each qubit from $\{\mu_i^{\pm}\}_{i\in[4]}$.\\
        Measure $\rho_{relax}$ in that basis and assign the binary variables according to the measurement result and basis for each qubit.\\
        Let the resulting solution be $x'$.\\
        Let $\mathrm{cut}(x)$ be the cut value of $x$.\\
        \If{$\mathrm{cut}(x)<\mathrm{cut}(x')$}{
            $x\leftarrow x'$
        }
    }
    \Return{x} 
\end{algorithm}
Unlike the case using the Pauli rounding algorithm, there is an approximation ratio bound for the MaxCut problem when using the magic state rounding algorithm.
It is obtained with the premise that the found relaxed state $\rho_{relax}$ has larger energy than the state associated with the optimal solution, i.e. $\mathrm{Tr}[H_{relax}\rho_{relax}]\geq\mathrm{Tr}[H_{relax}\rho_{opt}]=OPT$ where $\rho_{opt}$ is the quantum state which encodes the optimal solution using $(3,1)$-QRAC and $OPT$ is the optimal value of the instance.
\begin{theorem}[\cite{fuller2021approximate}]
    Given access to an oracle $\mathcal{O}_{relax}$ which prepares relaxed state $\rho_{relax}$ satisfying $\mathrm{Tr}[H_{relax}\rho_{relax}]\geq OPT$, the magic state rounding algorithm produces a solution to the MaxCut problem with expected approximation ratio $\mathbb{E}[\gamma]\geq\frac{5}{9}\approx0.555$.
    \label{thm:31ratio}
\end{theorem}
It was also proved by Fuller et al. that the approximation ratio bound for the QRAO using $(2,1)$-QRAC (which encodes up to two classical bits into a single-qubit) as $0.625$.
While the optimality of standard QAOA or VQE is often assumed when the obtained quantum state is the ground state, the above approximation ratios of QRAO do not require the exact ground state.
This is crucial because finding the exact ground state is known to be hard~\cite{kempe2006complexity}.

\section{The Role of Entanglement in Quantum Relaxation}
\begin{figure}[tb]
    \centering
    \includegraphics[scale=0.55]{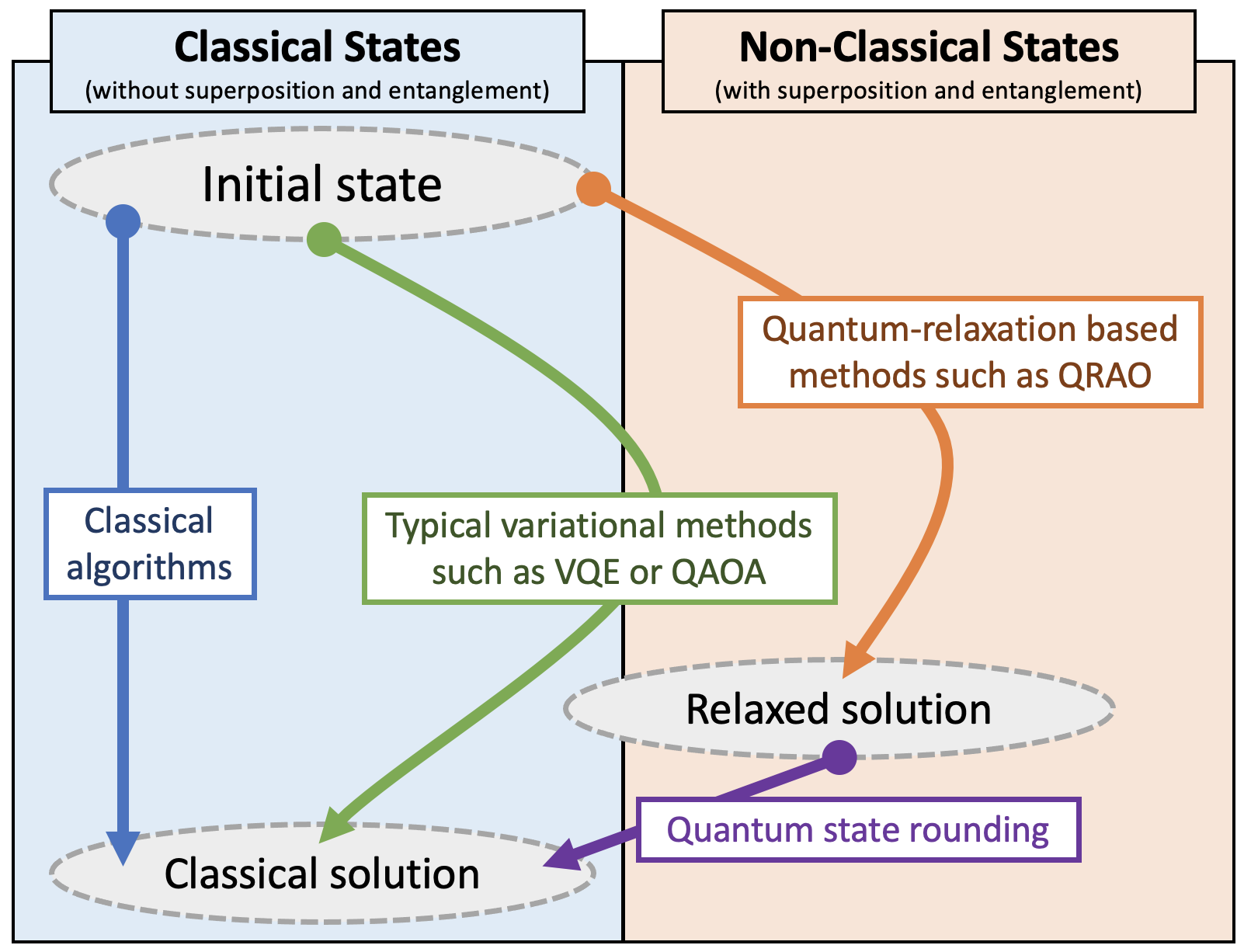}
    \caption{Illustration of the relation among classical algorithms, typical variational methods such as VQE or QAOA, and quantum-relaxation based methods such as QRAO}
    \label{fig:process_diff}
\end{figure}

First of all, we discuss the difference between typical classical-quantum hybrid quantum optimization algorithms such as VQE~\cite{peruzzo2014variational} or QAOA~\cite{farhi2014quantum} and quantum-relaxation based optimization algorithms such as QRAO~\cite{fuller2021approximate}.
As we see in the previous section, QRAO has the constant-factor (up to 3) space advantage against VQE or QAOA due to the QRACs which encode multiple classical bits into a single-qubit.
For the difference in the procedures, the typical approaches search for the problem's solution directly on variational methods while the quantum-relaxation based approaches explore the quantum state corresponding to the solution to the relaxed problem and extract the classical solution from the relaxed state by performing a quantum state rounding algorithm.
Then, the found quantum state is usually a classical state without superposition and entanglement in the former optimizers while the found relaxed states can be non-classical meaning that it contains superposition and entanglement in the latter approaches.
The illustration of the difference between them is visualized in \Cref{fig:process_diff}.
Because the classical states can be prepared even on classical computers, there is a possibility that the \textit{quantumness} is not needed in VQE or QAOA.
In other words, these algorithms may not be different from the classical optimization algorithms inherently.
In fact, there are some experimental results to support the above possibilities.
For VQE, there are experimental results that show no difference in solving various NP-hard problems in variational circuits with and without entanglement in the ansatz~\cite{nannicini2019performance,diez2021quantum}.
For QAOA, there are some problem instances for which constant-depth QAOA is worse than the classical counterpart~\cite{bravyi2020obstacles}.
This is a problem that is related to the very significance of solving optimization problems on a quantum computer.

On the other hand, non-classical states involve superposition and entanglement, and cannot be prepared on classical computers.
The problem Hamiltonian used in QRAO defined in \Cref{eq:H_relax} is not diagonal, and it contains non-classical states as the maximal eigenstates.
It implies that the found relaxed state in quantum relaxation may be non-classical and quantum-relaxation based optimizers inherently require \textit{quantumness}.
If it is the case, quantum relaxation not only can scale the instances of binary optimization problems solvable with limited quantum computers but also can benefit from quantum entanglement, unlike other quantum optimizers.
So, we hypothesize that quantum-relaxation based optimizers perform better in some cases with the help of quantumness.
In our experiment, we verified this hypothesis through intensive experiments on a quantum simulator.
Concretely, we concluded that there exist some problem instances for which quantum relaxation with entanglement layers in the ansatz finds the optimal solution while that without entanglement layers in the ansatz failed to find it.
Also, we change the various components of quantum relaxation and compare them to find some strategy when using quantum relaxation.
We will see more in detail about the methods or settings of the experiment in the following section.

\section{Experimental Settings}
\begin{figure}[tb]
    \centering
    \includegraphics[width=12cm]{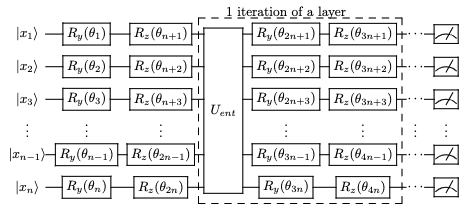}
    \caption{The quantum circuit diagram of the ansatz used in our experiments. The dotted lined part corresponds to a single iteration of a layer. The more layers, the more expressivity of the ansatz. Note that the single-qubit gates are Ry-Rz gates and $U_{ent}$ corresponds to the layer of $CNOT$ gates. The patterns of $U_{\mbox{ent}}$ used in our experiments are shown in \Cref{fig:ansatz}.}
    \label{fig:he_ansatz}
\end{figure}

In our experiment, we randomly generate $100$ instances of $3$-regular graphs with the number of the vertices ranging from $30$ to $48$ for the MaxCut problem.
We calculate relaxed Hamiltonians according to \Cref{eq:H_relax} and reverse the sign of each term.
After that, we searched for the ground state of the Hamiltonians on a simulator running VQE using a hardware-efficient ansatz~\cite{kandala2017hardware} whose layer consists of parametrized single-qubit Ry-Rz gates and entangling $CNOT$ gates (see \Cref{fig:he_ansatz}).
The \textit{depth} (more exactly the number of layers of single-qubit gates and entangling gates) of the VQE is changed from $0$ to $2$, with NFT (Nakanishi-Fuji-Todo) algorithm~\cite{nakanishi2020sequential} for up to 15 sweeps as an optimizer of the parametrized gates.
NFT algorithm is a gradient-free type optimizer and was designed for hardware-efficient ansatz.
A sweep corresponds to the single whole parameters' update in the algorithm.
We note that the NFT algorithm updates the parameter one by one while the general gradient-based optimizers usually update the whole parameters in one iteration.
Once we obtained the relaxed state, both the Pauli rounding algorithm and the magic state rounding algorithm are performed.
The number of measurements shown is set to be 1000.
The main purpose of the experiment is to compare the result with and without entanglement layers, in other words, the cases using ansatz with depth $0$ and depth $1$ or $2$.

\begin{figure}[tb]
    \begin{tabular}{cc}
        \begin{minipage}[b]{0.45\hsize}
            \centering
            \includegraphics[height=4.5cm]{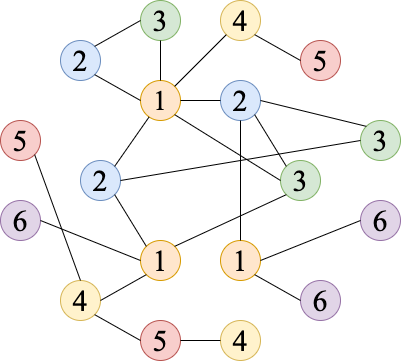}
            \subcaption{An example of a graph with 18 vertices. Each label of the vertices, ranging from 1 to 6, indicates the index of the qubit associated with it.}   
        \end{minipage} &
        \begin{minipage}[b]{0.45\hsize}
            \centering
            \includegraphics[height=4.5cm]{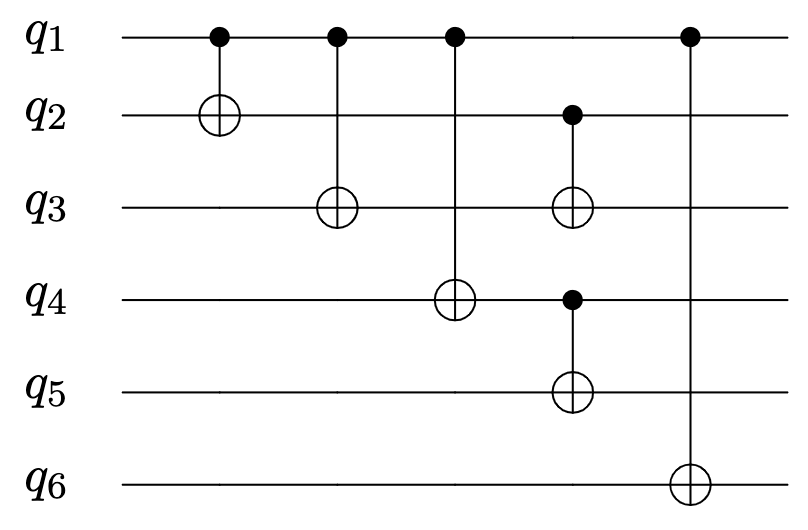}
            \subcaption{A compatible entanglement: $CNOT$ gates are applied between the qubits where there are edges between the vertices corresponding to each qubit.}
        \end{minipage} \\
        \begin{minipage}[b]{0.45\hsize}
            \centering
            \includegraphics[height=4.5cm]{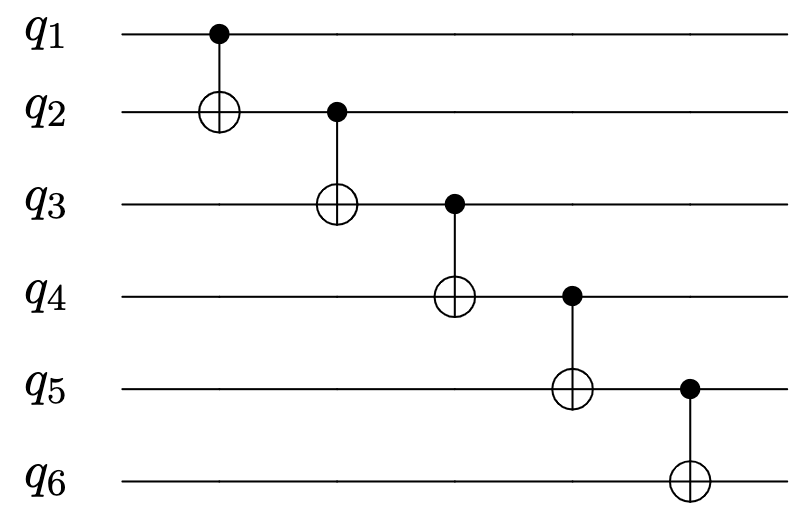}
            \subcaption{A linear entanglement: $CNOT$ gates are applied between the neighboring qubits $i$ and $i+1$.}       
        \end{minipage} &
        \begin{minipage}[b]{0.45\hsize}
            \centering
            \includegraphics[height=4.5cm]{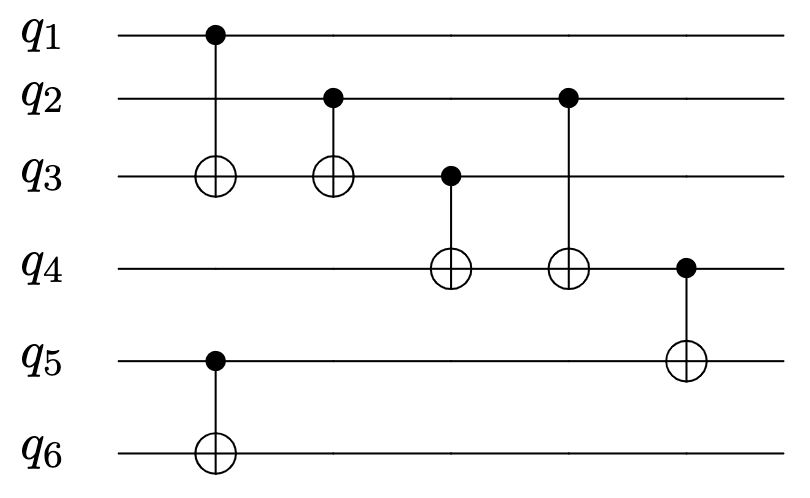}
            \subcaption{A random entanglement: $CNOT$ gates are applied to the randomly chosen pairs of qubits.}
        \end{minipage}
    \end{tabular}
    \caption{A graph and three types of entanglement $U_{\mbox{ent}}$ used in the quantum circuits of our experiments.}
    \label{fig:ansatz}
\end{figure}

Also, we modified the way to apply $CNOT$ gate in hardware-efficient ansatz and compared their performances.
Concretely three types of entanglement, i.e. \textit{compatible}, \textit{linear}, and \textit{random} are used like the previous research on VQE~\cite{diez2021quantum}.
\textit{Compatible} means that $CNOT$ gates are applied between the qubits of the encoding destinations of the vertices such that they are adjacent in the original graph before being encoded.
\textit{Linear} corresponds to the original hardware-efficient ansatz in that $CNOT$ gate is applied between qubits with adjacent indexes.
This is the most normal setting of the entanglement layers in hardware-efficient ansatz.
\textit{Random} indicates that $CNOT$ gates are applied to randomly chosen qubit pairs.
\Cref{fig:ansatz} shows an example of the difference between the entanglement layer forms of these three methods.
To examine the performance of the quantum-relaxation based optimizer for the MaxCut problem with possibly negative weights, we also performed the same experiments for the instances whose edge weights are $+1$ or $-1$ with the same probability as prior works for QAOA~\cite{farhi2014quantum}.

For algorithm implementation, we use the fast open-source quantum circuit simulator Qulacs~\cite{suzuki2021qulacs} and open-source quantum computing library Qiskit~\cite{aleksandrowicz2019qiskit}.
We also used some parts of the public implementation of QRAO\footnote{\url{https://github.com/qiskit-community/prototype-qrao}}.
We did these experiments on the Wisteria supercomputer system at Supercomputing Division, Information Technology Center, The University of Tokyo\footnote{\url{https://www.cc.u-tokyo.ac.jp/en/supercomputer/wisteria/system.php}}.
Especially, the `regular-a` resource group (1-8 nodes, 448GiB memory) in the Wisteria-A (Aquarius) system and the `prepost` resource group (1 node, 340GiB memory) in Wisteria/BDEC-01 system are used in the experiment.
Wisteria-A system's CPU has 2 processors (Intel Xeon Platinum 8360Y, 36 cores) and 512GiB memory.

\section{Results and Discussions}
\subsection{Unweighted Instances}
To compare the result among the case of various members of nodes, we use the relaxed state energy value divided by the optimal MaxCut value (we call it \textit{normalized relaxed energy} for simplicity) to evaluate the performance of the step to explore the relaxed state and use the approximation ratio to evaluate the performance of the quantum-relaxation based optimization algorithm.
We note that the value of the normalized relaxed energy should be larger than $1$ to satisfy the assumption in the proof of the approximation ratio in \Cref{thm:31ratio}.
We will see the individual results at first and summarize them later.
\begin{figure}[tb]
    \begin{tabular}{cc}
        \begin{minipage}[t]{0.45\hsize}
            \centering
            \includegraphics[height=4cm]{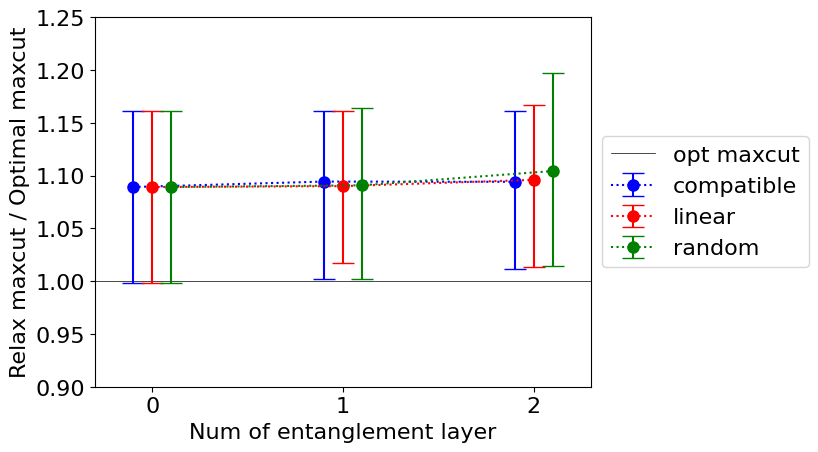}
            \subcaption{The average normalized relaxed energy and the number of entanglement layers ($30$ nodes)}
            \label{fig:relax_30}
        \end{minipage} &
        \begin{minipage}[t]{0.49\hsize}
            \centering
            \includegraphics[height=4cm]{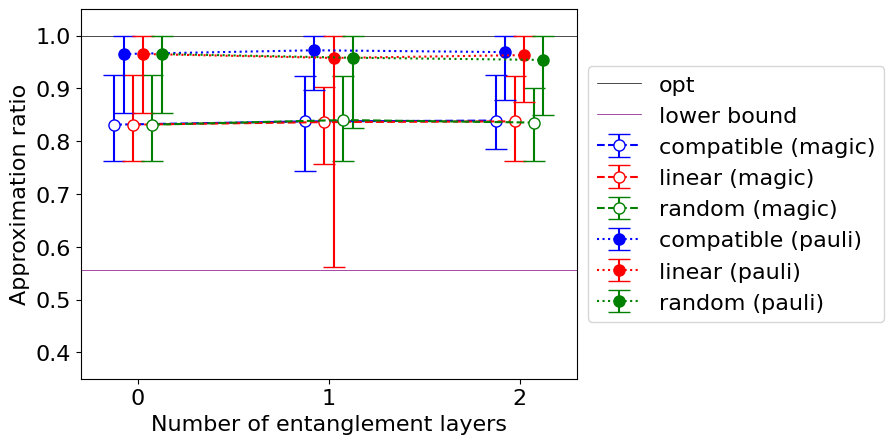}
            \subcaption{The average approximation ratio and the number of entanglement layers ($30$ nodes)}
            \label{fig:round_30}
        \end{minipage}
    \end{tabular}
    \caption{The experimental result in the case of $30$ nodes unweighted MaxCut instances}
    \label{fig:result_30}
\end{figure}
\Cref{fig:result_30} shows the result in the case of $30$ nodes.

\Cref{fig:relax_30} shows the relationship between the normalized relaxed energies and the number of entanglement layers in ansatz.
The vertical axis represents the average normalized relaxed energy of the found relaxed states for $100$ instances and the horizontal axis represents the number of entanglement layers.
$0$ layer means that the ansatz without entanglement layer is used in the algorithm.
The color of the plots represents the type of entanglement layer forms.
A compatible entanglement, a linear entanglement, and a random entanglement are represented in blue, red, and green respectively.
The error bar of the plot represents the maximum and minimum normalized relaxed energy among the $100$ instances.
The black horizontal line at the normalized relaxed energy $1.0$ express the border of the premise of the approximation ratio given by \Cref{thm:31ratio}.
For all three entanglement types and for all $100$ instances, the found relaxed state's normalized energy exceeds the border.
On average, the value of the normalized relaxed energy becomes slightly larger as the number of entanglement layers increases.
It implies that introducing quantum entanglement in the ansatz contributes to the search for the entangled state with higher energy to some extent.

\Cref{fig:round_30} shows the result of the quantum state rounding from the relaxed state.
The vertical axis represents the approximation ratio of the decoded classical solution's value.
The horizontal axis represents the number of entanglement layers.
The type of the entanglement is expressed as the same colors of the point as in \Cref{fig:relax_30}.
The white-out points correspond to the average results of the magic state rounding algorithm described in \Cref{alg:magic}, and the filled points express the average results of the Pauli rounding algorithm described in \Cref{alg:pauli}.
As we see in the previous section, there is no guarantee that the Pauli rounding algorithm performs well for the highly-entangled relaxed state while the magic state rounding algorithm has the theoretical bound $0.555$ of the approximation ratio if the normalized relaxed energy exceeds $1.0$.
In \Cref{fig:round_30}, the approximation ratio is larger than the theoretical bound.
Furthermore, the Pauli rounding algorithm performs better than the magic state rounding algorithm for almost all instances and the number of entanglement layers.
Though the magic state rounding algorithm fails to find an optimal solution for all $100$ instances, The Pauli rounding algorithm succeeds in getting an optimal solution for some instances.
These facts imply that the relaxed state may not be very entangled practically.
Also, though it may be because the problem instance size involved in our experiment is small ($\leq48$ nodes), even in the case without entanglement layers, we can find optimal solutions for some instances.
The average approximation ratio is not either improved significantly in the case of $30$ nodes.
\begin{figure}[tb]
    \begin{tabular}{cc}
        \begin{minipage}[t]{0.45\hsize}
            \centering
            \includegraphics[height=4cm]{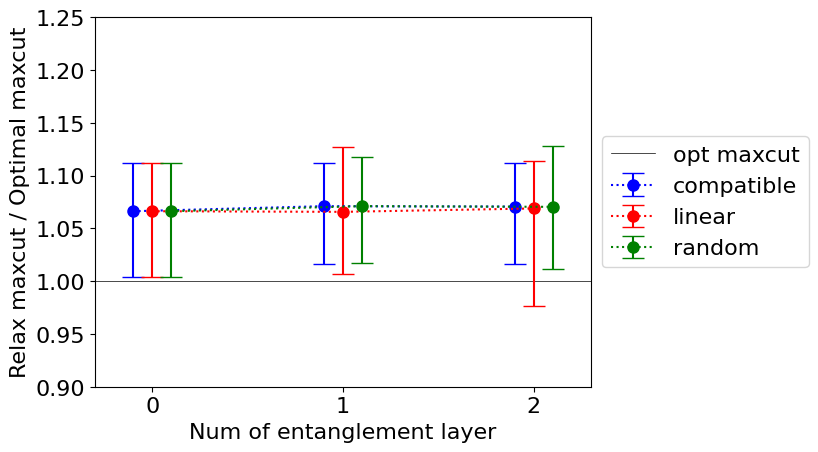}
            \subcaption{The average normalized relaxed energy and the number of entanglement layers ($46$ nodes)}
            \label{fig:relax_46}
        \end{minipage} &
        \begin{minipage}[t]{0.49\hsize}
            \centering
            \includegraphics[height=4cm]{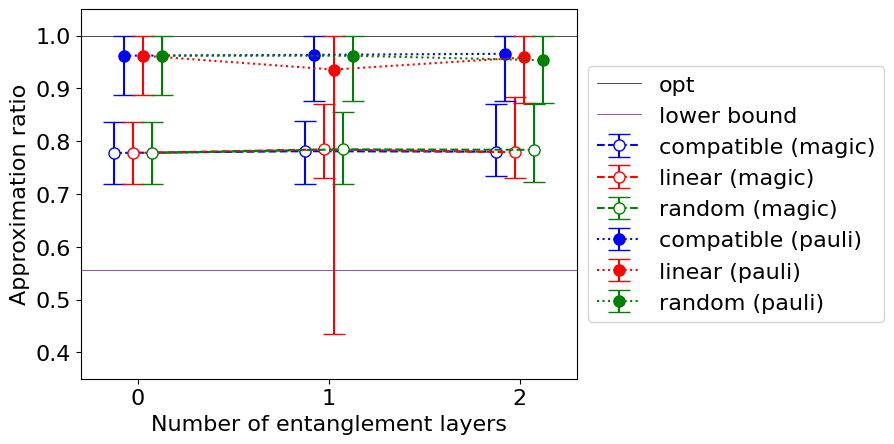}
            \subcaption{The average approximation ratio and the number of entanglement layers ($46$ nodes)}
            \label{fig:round_46}
        \end{minipage}
    \end{tabular}
    \caption{The experimental result in the case of $46$ nodes unweighted MaxCut instances}
    \label{fig:result_46}
\end{figure}

Let us see the results for larger graph instances.
\Cref{fig:result_46} shows the result for $46$ nodes instances.
The number of entanglement layers seems to have no effect on the normalized relaxed energy in the case of $46$ nodes.
The different point from the case of $30$ nodes lies in the result of the Pauli rounding.
Though the normalized relaxed energy is almost the same, there is a bad graph instance where the Pauli rounding's performance is quite bad when the type of entanglement form is linear and the number of entanglement layers is $1$.
We focus on this bad instance.
\begin{figure}[tb]
    \begin{tabular}{cc}
        \begin{minipage}[t]{0.45\hsize}
            \centering
            \includegraphics[height=4cm]{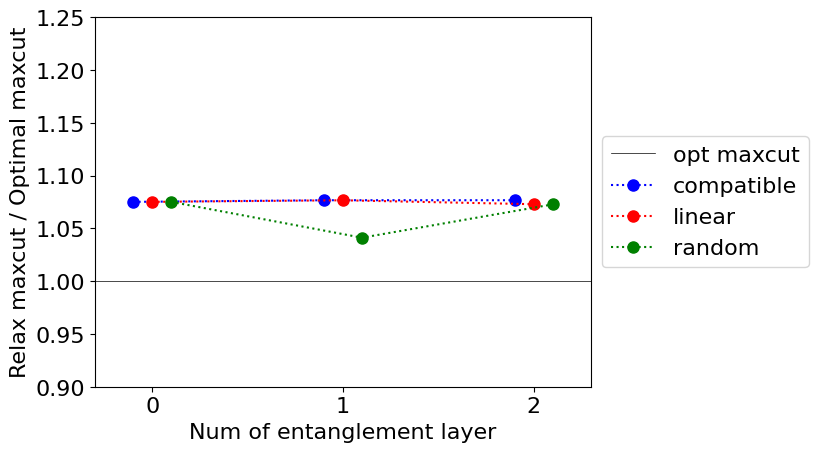}
            \subcaption{The normalized relaxed energy and the number of entanglement layers for the bad instance with $46$ nodes}
            \label{fig:relax_bad}
        \end{minipage} &
        \begin{minipage}[t]{0.49\hsize}
            \centering
            \includegraphics[height=4cm]{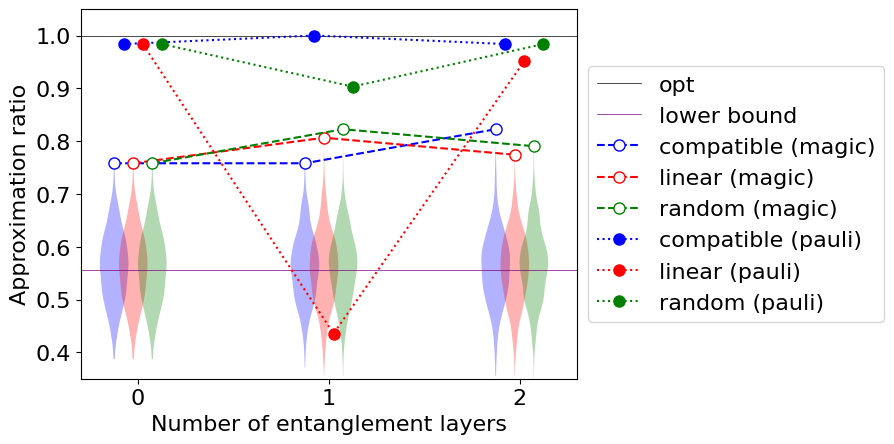}
            \subcaption{The approximation ratio and the number of entanglement layers for the bad instance with $46$ nodes}
            \label{fig:round_bad}
        \end{minipage}
    \end{tabular}
    \caption{The experimental result for the bad instance with $46$ nodes}
    \label{fig:result_bad}
\end{figure}

\Cref{fig:result_bad} shows the results for the bad $46$ nodes instance mentioned above.
The violin plot in \Cref{fig:round_bad} represents the distribution of the decoding results of the $1000$ shots (iterations) of the magic state rounding algorithm.
The best one of them is output as the classical solution (see \Cref{alg:magic}).
The interesting point is that the normalized relaxed energy values of the linear entanglement are almost the same in each number of entanglement layers while the result of the Pauli rounding drastically becomes worse than $\frac{1}{2}$ in the case of $1$ entanglement layer and again becomes almost optimal in the case of $2$ entanglement layers.
Conversely, the performance of the magic state rounding algorithm becomes better in the case of a single entanglement layer.
These facts imply that the linear entanglement type ansatz prepares the very entangled state in the case of the single entanglement layer.
The same patterns (the Pauli rounding's performance is quite bad when the number of entanglement layers is $1$ or $2$) also appear in the case of $32$, $34$, $40$, and $44$ nodes strongly meaning that the approximation ratio is less than half.
However, all of these patterns only appear in the linear entanglement type ansatz.
\begin{figure}[tb]
    \begin{tabular}{cc}
        \begin{minipage}[t]{0.45\hsize}
            \centering
            \includegraphics[height=4cm]{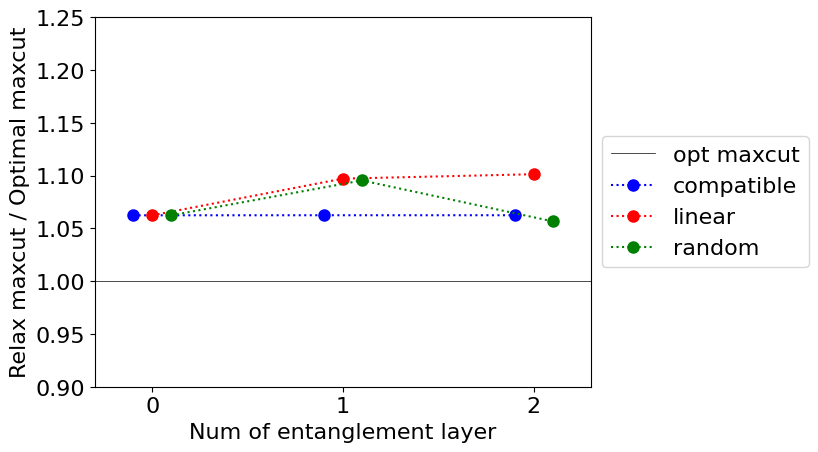}
            \subcaption{The normalized relaxed energy and the number of entanglement layers for the good instance with $46$ nodes}
            \label{fig:relax_good}
        \end{minipage} &
        \begin{minipage}[t]{0.49\hsize}
            \centering
            \includegraphics[height=4cm]{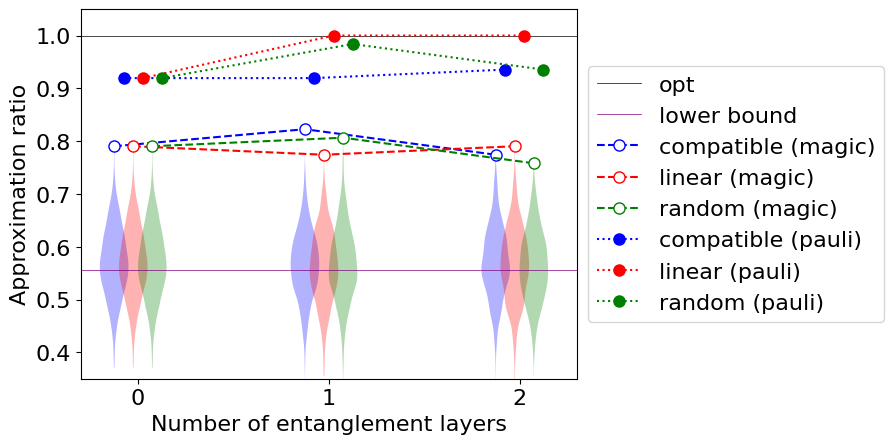}
            \subcaption{The approximation ratio and the number of entanglement layers for the good instance with $46$ nodes}
            \label{fig:round_good}
        \end{minipage}
    \end{tabular}
    \caption{The experimental result for the good instance with $46$ nodes}
    \label{fig:result_good}
\end{figure}

On the other hand, there exist graph instances for which the quantum relaxation fails to find an optimal solution without entanglement layers and can find it with entanglement layers.
\Cref{fig:result_good} shows the result for the good instance for which quantum relaxation using the Pauli rounding algorithm fails to find an optimal solution without entanglement layers but succeeds in finding it with linear entanglement layers.

In summary, we have the following observations:
\begin{itemize}
    \item Practically, the Pauli rounding algorithm usually outperforms the magic rounding algorithm.
    \item The average approximation ratio of the Pauli rounding algorithms is around $0.85$-$1.0$ and that of the magic state rounding algorithm is around $0.7$-$0.9$.
    \item The assumption of the proof of the approximation ratio is satisfied in almost all the cases and there is no instance that violates the approximation ratio bound. In other words, the result of the magic state rounding algorithm is always better than $\frac{5}{9}\approx0.555$.
    \item On the other hand, the Pauli rounding algorithm's performance sometimes becomes quite worse (and the resulting approximation ratio is less than half). This phenomenon only happens in the case using the linear entanglement type ansatz.
    \item Higher normalized relaxed energy does not always generate a better solution.
\end{itemize}
Let us go back to the main purpose of our experiments, which is to verify experimentally that the quantum relaxation performs better in some cases if we have entanglement layers in ansatz.
On average, there seems to be almost no effect of the increase in the number of entanglement layers on the approximation ratio.
However, there certainly exists some instances such as the instance shown in \Cref{fig:result_good} where the performance of the quantum relaxation is improved by introducing entanglement layers and even optimal solution is found.

\begin{table}[tb]
    \centering
    \caption{The number of good instances among $100$ instances for each number of nodes where quantum relaxation succeeded in finding an optimal solution. $L=0$ corresponds to the case without entanglement and $L\leq n$ means the number of instances where optimal solutions were found in at least one of $L=i\ (i\in\{0,...,n\})$.}
    \begin{tabular}{|c|rrrrrrrrrr|}
        \hline
        num. of nodes & 30 & 32 & 34 & 36 & 38 & 40 & 42 & 44 & 46 & 48 \\ \hline
        ($L=0$) & 34 & 16 & 17 & 15 & 13 & 17 & 11 & 13 & 11 & 15 \\ \hline
        compatible ($L\leq1$) & \textbf{43} & 26 & 24 & 29 & 21 & 27 & 18 & 21 & 18 & 24 \\
        linear ($L\leq1$) & \textbf{43} & 29 & \textbf{30} & \textbf{30} & 23 & 24 & 17 & \textbf{27} & 21 & \textbf{25} \\
        random ($L\leq1$) & \textbf{43} & \textbf{36} & 23 & 27 & \textbf{24} & \textbf{32} & \textbf{24} & 21 & \textbf{26} & 21 \\ \hline
        compatible ($L\leq2$) & 46 & 32 & 31 & 36 & 26 & 34 & 24 & 23 & 21 & 25 \\
        linear ($L\leq2$) & \textbf{52} & 37 & \textbf{41} & \textbf{42} & 31 & 36 & 28 & \textbf{36} & \textbf{31} & \textbf{32} \\
        random ($L\leq2$) & 47 & \textbf{48} & 35 & \textbf{42} & \textbf{37} & \textbf{38} & \textbf{35} & 26 & \textbf{31} & 26 \\ \hline
    \end{tabular}
    \label{tab:opt}
\end{table}

Let $L$ be the number of entanglement layers in ansatz.
\Cref{tab:opt} enumerates the number of the instances among $100$ instances for each number of nodes where the quantum relaxation succeeded in finding an optimal solution without entanglement layers ($L=0$) and the total number of the instances where optimal solutions were found in at least one of the cases of $L=i\ (i=0,1)$ and $L=i\ (i=0,1,2)$.
By comparing $L=0$, $L\leq1$, and $L\leq2$, we conclude that the coverage of the instances for which the quantum relaxation can find an optimal solution is spread if we run the algorithm with entanglement layers.
In other words, we observed that \textit{quantumness} enhances the quantum relaxation's performance for some instances.

From the above discussion, the good practical strategy is to increase the number of the entanglement layers up to some threshold $L_{max}$ and output the best solution.
The result of our experiment implies that $L_{max}=2$ is better than $L_{max}=0$ in this strategy.
If we have the time limit of the computation, increasing the number of the entanglement layers in the middle seems to be better than solving without entanglement layers the whole time from our results because the distributions of the problem instances for which the quantum relaxation is good at finding an optimal solution are different in the cases with and without entanglement layers in our experiment.

Finally, we discuss the difference between the type of entanglement forms.
We note that the number of $CNOT$ gates that appeared at the random entanglement type ansatz and the number of $CNOT$ gates in the compatible entanglement type ansatz are equal.
The number of $CNOT$ gates in the linear entanglement type ansatz is equal to the number of the qubit minus $1$ and is smaller than that of the compatible and the random type entanglement layers.
The bold number in the \Cref{tab:opt} designates the maximum number of the instances whose optimal solutions were found by quantum relaxation among compatible, linear, and random entanglement types with the common condition $L\leq i\ (i=1,2)$.
For both $L\leq1$ and $L\leq2$, compatible entanglement is worse than the other two.
The fact that random entanglement is better than compatible entanglement implies that it is better to neglect the structure of the instance graph.
It may be because reflecting the graph structure increases the probability that the optimizer finds a local optimum solution.
Considering the implementation costs or the noise on the real devices, using linear entanglement types seems to be the best strategy because the number of CNOT gates used in random entanglement is $O(|E(G)|)$ while that used in linear entanglement is $O(|V(G)|)$.

\subsection{Weighted (+1 or -1) Instances}
In this section, we will show the result of the same experiments for weighted instances whose weights are $+1$ or $-1$ and are assigned randomly.
For weighted instances, the quantum relaxation has no approximation ratio bound.
Actually, the magic state rounding algorithm outputs the solution whose approximation ratio is less than $\frac{5}{9}\approx0.555$ for many cases.
On the other hand, in our experiment, quantum relaxation finds many optimal solutions to various instances when using the Pauli rounding.

\begin{table}[tb]
    \centering
    \caption{The number of good weighted instances among $100$ instances for each number of nodes where quantum relaxation succeeded in finding an optimal solution. $L=0$ corresponds to the case without entanglement and $L\leq n$ means the number of instances where optimal solutions were found in at least one of $L=i\ (i\in\{0,...,n\})$.}
    \begin{tabular}{|c|rrrrrrrrrr|}
        \hline
        num. of nodes & 30 & 32 & 34 & 36 & 38 & 40 & 42 & 44 & 46 & 48 \\ \hline
        ($L=0$) & 29 & 21 & 12 & 17 & 21 & 12 & 14 & 9 & 16 & 11 \\ \hline
        compatible ($L\leq1$) & 43 & 31 & 20 & 26 & 31 & 21 & \textbf{26} & \textbf{19} & \textbf{26} & 18 \\
        linear ($L\leq1$) & 41 & 29 & \textbf{24} & 24 & \textbf{32} & 20 & 22 & 16 & 23 & \textbf{20} \\
        random ($L\leq1$) & \textbf{55} & \textbf{40} & 21 & \textbf{29} & 28 & \textbf{26} & 24 & 16 & 24 & \textbf{20} \\ \hline
        compatible ($L\leq2$) & 51 & 33 & 28 & \textbf{34} & 37 & 26 & \textbf{29} & 21 & 31 & \textbf{31} \\
        linear ($L\leq2$) & 51 & 39 & \textbf{39} & 33 & \textbf{38} & 27 & 28 & \textbf{28} & \textbf{33} & 24 \\
        random ($L\leq2$) & \textbf{58} & \textbf{47} & 32 & 32 & 36 & \textbf{33} & 28 & 20 & 25 & 29 \\ \hline
    \end{tabular}
    \label{tab:opt_w}
\end{table}

\Cref{tab:opt_w} enumerates the number of the weighted instances among $100$ instances for each number of nodes where the quantum relaxation succeeded in finding an optimal solution without entanglement layers ($L=0$) and the total number of the instances where optimal solutions were found in at least one of the cases of $L=0$, $L=1$, and $L=2$.
By comparing $L=0$ and $L\leq2$, we can conclude that \textit{quantumness} also enhances the quantum relaxation's performance for some instances experimentally even for the weighted ($+1$ or $-1$) MaxCut problem.
The bold number in the \Cref{tab:opt_w} designates the maximum number of the instances whose optimal solutions were found by quantum relaxation among compatible, linear, and random entanglement types with the common condition $L\leq i\ (i=1,2)$.
Unlike the unweighted case, compatible entanglement outperforms the other two in some cases.
Again considering the implementation cost, using linear entanglement seems to be better.

\section{Conclusion and Future Directions}
In this paper, we experimentally analyze the quantum-relaxation based optimization algorithm represented by QRAO~\cite{fuller2021approximate} and experimentally check that the quantumness (i.e. quantum entanglement) increases the performance of the quantum-relaxation based optimizers.
Concretely, we run the $(3,1)$-QRAO for various both unweighted and weighted ($+1$ or $-1$) $3$-regular MaxCut problem instances and compare the results between the case using the ansatz without entanglement layers and the case using the ansatz with $1$ or $2$ entanglement layers.
In conclusion, we observed that there exist some instances for which the quantum relaxation without entanglement layers in ansatz fails to find an optimal solution while that with entanglement layers in ansatz succeeds.

There are some future directions.
In our experiment, we use one of the gradient-free optimizers, NFT algorithm~\cite{nakanishi2020sequential,ostaszewski2021}.
Though the algorithm converges quite faster than the general gradient-based optimizers (and actually makes it possible to run the simulation for the instances up to $48$ vertices), there are even better optimizers designed for the ansatz made of free-axis (Fraxis) gates~\cite{watanabe2021optimizing}, or free quaternion selection (FQS) gates~\cite{wada2022,wada2022full}.  
While we optimize the angle of the rotation gates in the hardware-efficient ansatz, their algorithm optimizes the direction of the axis of the free-axis gates which converges faster and even finds a better eigenstate of the Hamiltonian.
The speedup of the convergence may enable us to run the experiment of the quantum relaxation on larger instances. However, at the moment basically the bottleneck of the running time of the experiment is due to the number of qubits to be simulated.

The other possible direction is considering a warm-starting.
Warm-starting is a technique to start the search for the ground state of the problem Hamiltonian from a good initial solution and is applied to standard quantum optimizers such as QAOA~\cite{tate2020bridging,egger2021warm}.
The candidates of the initial solution are the relaxed solution to the SDP of the MaxCut problem or the approximate classical solution obtained by Goemans-Williamson's SDP rounding algorithm~\cite{goemans1995improved}.
By using the rotation gates, we can encode the relaxed (non-integral) solution into the quantum circuit.
We consider using the result of the Pauli rounding result as the candidate of the good initial solution and encode it by using the corresponding QRAC as the initial state in the variational methods.
It is likely that by using this good initial solution, the result of the magic state rounding algorithm is improved like warm-starting for standard quantum optimizers.
However, though it may be because the problem instance is too small to evaluate the performance of the warm-starting, the result is not improved at all for the candidate instances in our experimental analysis.
It seems that we need more techniques to use  warm-starting for QRAO.

Also, we discuss the construction of the recursive algorithm for quantum relaxations corresponding to recursive QAOA~\cite{bravyi2020obstacles}.
Recursive QAOA is the algorithm to fix the parity of the variables of the most correlated edge with the problem Hamiltonian, and reduce $n$-vertices MaxCut problem instance to the instance of the size $n-1$.
Then, we recursively call the algorithm until the size of the problem becomes less than the threshold $n_c$.
Bravyi et al. confirmed by experiment that $p$-level QAOA achieves the approximation ratio at most $\frac{p}{p+1}$ while the $1$-level recursive QAOA achieves the approximation ratio $1$ where the level means the number of the repetition of the layer in the variational circuits.
It implies that recursive QAOA not only makes the necessary quantum circuit size smaller but also improves the approximation ratio of the QAOA.
Our motivation is to consider the same kind of recursive algorithm for the quantum-relaxation based optimizers too.
Naively, we can calculate the correlation of each edge by measuring the qubits associated with its endpoints in the corresponding ($X$, $Y$, or $Z$) basis.
Then, we can reduce the number of vertices one by one for each recursion.
However, because the quantum-relaxation based approaches encode multiple vertices into a smaller number of qubits, the number of the qubits may not be reduced for each recursion (note that we may overcome this by encoding two bits of information in a qubit as explained in~\cite{teramoto2023quantum}).
Even worse, it may be possible that the number of qubits needed is increased because the density of the graph instance becomes larger for each recursion generally.
This is because QRAO has the constraint that the endpoints of the arbitrary edge of the graph must be associated with different qubits.
So, we consider the smarter way of recursion than the above naive methods.
The idea is to fix the relationship between the pair of three bits encoded into the most correlated two qubits, and reduce the three vertices encoded in one of the qubits at once.
The method is that for each pair of two qubits $(i,j)$, we evaluate the value
\begin{equation}
    \mathrm{Tr}[(\rho_{x_1,x_2,x_3})_i\otimes(\rho_{x_1',x_2',x_3'})_j \rho_{relax}]
    \label{eq:cor_qubit_pair}
\end{equation}
for every pair $x_1,x_2,x_3,x_1',x_2',x_3'\in\{0,1\}^6$ and choose the pair of qubits $(i,j)$ and the pairs $x_1,x_2,x_3,x_1',x_2',x_3'$ where the sign of \Cref{eq:cor_qubit_pair} is maximized.
This is the definition of the most correlated two qubits.
The evaluation of values in \Cref{eq:cor_qubit_pair} can simultaneously be done by only a few shots of measurements (logarithmic number of the terms we want to approximate) by using a technique so-called \textit{classical shadow}~\cite{cotler2020quantum,huang2020predicting,hadfield2022measurements}.
To fix the relation between $x_1,x_2,x_3$ and $x_1',x_2',x_3'$, we have to define the $3\times 3$ transformation matrix from one to the other.
It may be possible to improve the approximation ratio of the original quantum relaxation like recursive QAOA~\cite{bravyi2020obstacles}.

Another idea is to have quantum relaxations with different types of formulations.
QRAO~\cite{fuller2021approximate} is formulated based on the binary optimization form in \Cref{eq:MaxCut} while there exist quantum optimizers using a formulation based on graph Laplacian matrices to obtain exponentially qubit-saving quantum optimizers~\cite{ranvcic2021exponentially,winderl2022comparative,chatterjee2023solving}.
It may be possible to consider the relaxed version of these optimizers using only the logarithmic number of qubits to the size of the graph instance.
It seems interesting to evaluate the theoretical performance guarantee of such optimizers and compare them with the ones of the \textit{QRAO-type} quantum relaxations using quantum random access codes.

Finally, we recently obtained a result that quantum entanglement helps improve the approximation ratio~\cite{teramoto2023quantum} due to the use of $(3,2)$-QRAC~\cite{imamichi2018constructions}, i.e., the QRAC that uses two-qubit entanglement to encode/decode bits into qubits.
This suggests that it may be possible to improve the approximation ratio by considering QRACs on more qubits with higher success probabilities.
\bibliographystyle{plain}
\bibliography{references}
\end{document}